\documentclass[journal]{IEEEtran}
\ifCLASSINFOpdf
   \usepackage[pdftex]{graphicx}
\else
\fi
\usepackage{amsmath}

\usepackage{breqn}
\usepackage{xcolor}
\usepackage{graphicx} 
\usepackage{caption,subcaption}
\usepackage{mathtools}
\usepackage{booktabs}
\usepackage{algorithm}
\usepackage{algpseudocode}
\usepackage{multirow}
\usepackage{array}

\usepackage{mathtools}
\usepackage{amsfonts}
\usepackage{booktabs}
\usepackage{algorithm}
\usepackage{algpseudocode}
\usepackage{amsmath}

\begin{document}
%

\title{Side-aware Meta-Learning for Cross-Dataset Listener Diagnosis with Subjective Tinnitus}

\author{Yun~Li, Zhe~Liu, 
        Lina~Yao,
        Molly Lucas,
        Jessica~J.M.Monaghan, 
        and Yu Zhang
\thanks{Yun Li, Zhe Liu, and Lina Yao are with the School of Computer Science and Engineering, University of New South Wales. Australia e-mails: yun.li5@unsw.edu.au, zheliu912@gmail.com, lina.yao@unsw.edu.au.}
\thanks{Molly Lucas is with Department of Psychiatry and Behavioral Sciences, Stanford University, Palo Alto, CA 94305, USA. e-mail: molly.v.lucas@gmail.com}
\thanks{Jessica J.M.Monaghan is with National Acoustic Laboratories, The Australian Hearing Hub, Sydney, NSW, 2109, Australia
Macquarie University, The Australian Hearing Hub, NSW, 2109, Sydney, Australia. Australia e-mail: jessica.monaghan@gmail.com.}
\thanks{Yu Zhang is with the Department of Bioengineering, Lehigh University, Bethlehem, PA 18015, USA. e-mail: yuzi20@lehigh.edu}
}


%



\maketitle

\begin{abstract}
With the development of digital technology, machine learning has paved the way for the next generation of tinnitus diagnoses. Although machine learning has been widely applied in EEG-based tinnitus analysis, most current models are dataset-specific. 
Each dataset may be limited to a specific range of symptoms, overall disease severity, and demographic attributes; further, dataset formats may differ, impacting model performance. 
This paper proposes a side-aware meta-learning for cross-dataset tinnitus diagnosis, which can effectively classify tinnitus in subjects of divergent ages and genders from different data collection processes. Owing to the superiority of meta-learning, our method does not rely on large-scale datasets like conventional deep learning models. Moreover, we design a subject-specific training process to assist the model in fitting the data pattern of different patients or healthy people. Our method achieves a high accuracy of 73.8\% in the cross-dataset classification. We conduct an extensive analysis to show the effectiveness of side information of ears in enhancing model performance and side-aware meta-learning in improving the quality of the learned features.

\end{abstract}
\begin{IEEEkeywords}
Electroencephalography, subject-independent, cross-dataset, meta-learning, tinnitus.
\end{IEEEkeywords}

%
\IEEEpeerreviewmaketitle

\section{Introduction}
With the development of digital technologies, digital processing has been the core point of the next generation of tinnitus therapy~\cite{searchfield2021state}. State-of-the-art digital processing relies on the improvement of artificial intelligence to physiological sensors, which enables the personalization of the patient-centered therapies~\cite{hunn2016market,crum2019hearazbles}. The most representative and commonly used artificial technology applied to improve tinnitus therapies is machine learning~\cite{koprinkova2015artificial}. For example, machine learning has been widely applied in the analysis of electroencephalogram (EEG)~\cite{allgaier2021deep}, auditory brainstem response (ABR)~\cite{liu2021generalizable}, and functional magnetic resonance imaging (fMRI)~\cite{shahsavarani2020comparing}. In particular, EEG can be an effective and inexpensive data source to analyze the neural feedback of tinnitus patients~\cite{sun2019multi}.

Extensive EEG-based machine learning research~\cite{sun2019multi,liu2021generalizable,li2016svm,saeidi2021neural,emami2017eeg} has shown that machine learning is able to discover differences in signals between tinnitus patients and healthy people by learning patterns from given (i.e., training) datasets. Therefore, EEG dataset can be a reliable data source that enables machine learning models to assist in diagnosing tinnitus. While some research~\cite{quiet-svm-tinnitus,basic-tinnitus-subject-dependent,zscoretinnitus} focuses on learning the pattern difference of the existing tinnitus and control subjects, there exist significant variance between subjects~\cite{liu2021generalizable}. When between-subject variance is high, these models may only work when the testing signals are sampled from known subjects or if their distributions closely mirror those seen within the training dataset. To enhance the model robustness, especially with regards to handling new subjects, some other research~\cite{liu2021generalizable,power_value} enables the model to be aware of the subject variance. By understanding the subject variance in signals, models can mitigate the corresponding negative influences in the prediction.  However, in real-world scenarios, many EEG datasets~\cite{liu2021generalizable,sun2019multi,allgaier2021deep} may collect data using different experimental environments and thus build data with diverse formats. The different data formats may cause an extremely significant domain shift. Therefore the majority of models are still limited to a single dataset and may not be able to be generalized across distinct datasets.

\begin{figure}
\centering 
\begin{subfigure}{0.24\textwidth}
  \includegraphics[width=\textwidth]{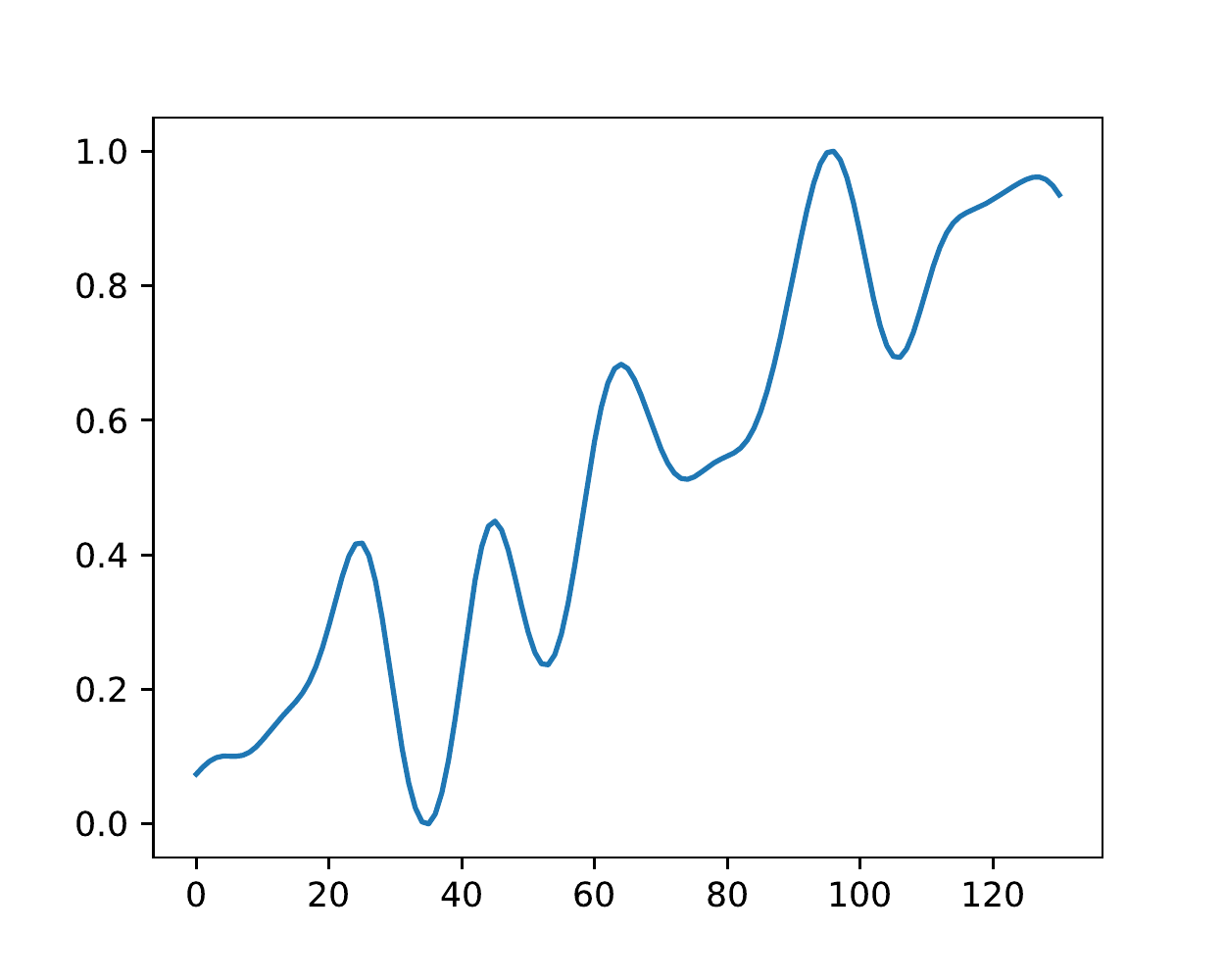}
    \centering
  \caption{Left side.}
\end{subfigure}\hfil 
\begin{subfigure}{0.24\textwidth}
  \includegraphics[width=\textwidth]{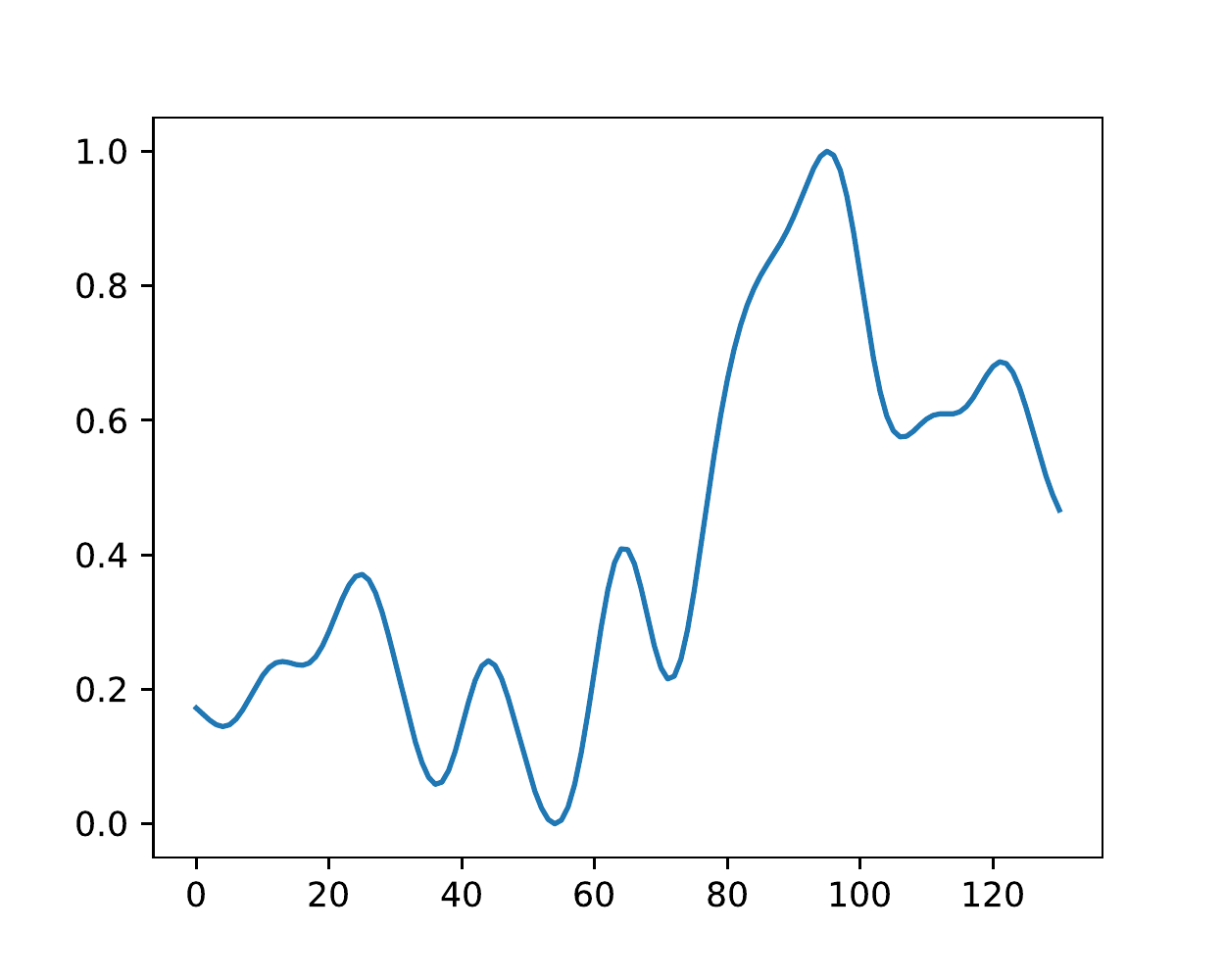}
    \centering
  \caption{Right side.}
\end{subfigure}\hfil 
    \caption{Illustration of signal difference on (a) the left and (b) the right sides of the same subject.}\label{left_right_dif}
\end{figure}

In the other related fields, it has been shown that the cross-dataset domain shift will heavily impair model performance in classifying EEG signals~\cite{cimtay2020investigating,xu2020cross,lan2018domain,lin2019constructing,rayatdoost2018cross}. For example, unsupervised domain adaptation~\cite{lan2018domain} and transfer learning~\cite{lin2019constructing} show good performance on cross-dataset EEG-based emotion classification. However, these methods rely on large-scale datasets like common deep learning methods. Meanwhile, some advanced technologies in machine learning can be applied to help tackle the cross-dataset domain shift in EEG. One such technique, model-agnostic meta-learning has been widely applied in computer vision~\cite{finn2017model}, natural language processing~\cite{obamuyide2019model}, and recommender systems~\cite{bharadhwaj2019meta}. The model-agnostic meta-learning is compatible with any traditional machine learning method and does not rely on large-scale datasets. It can learn a set of generalized basic model parameters based on existing datasets and then be quickly adapted to new datasets by fitting a few samples~\cite{hospedalesmeta}, which effectively prevents domain shift among different datasets. However, directly using these methods may not fully capture the characteristic information of tinnitus signals. As shown in Figs.~\ref{left_right_dif} (a-b), there exists a difference between the EEG signals corresponding to sounds presented to the left and right ears. Traditional methods may confuse this side difference with the difference in subject variance. This may lead models to learn false information that these signals are from two different subjects, which will impair model performance and cause inaccurate conclusions in the further analysis of the learned representation.

In this paper, we propose a domain-specific meta-learning, named Side-aware Meta-learning (SMeta), for tinnitus diagnosis. SMeta utilizes the subject difference of left/right ears and takes the difference as an auxiliary indicator to assist meta-learning. To conduct subject-specific meta-learning, we also propose subject-independent task training for SMeta, which allows episode-wise subject-specific training. We further implement SMeta based on autoencoder and design two model variants, i.e., Side-aware Meta-AutoEncoder (SMeta-AE) and Side-aware Meta-Siamese-AutoEncoder (SMeta-SAE). Both variants outperform state-of-the-art methods.  Our major contributions in this paper can be summarized as follows:
\begin{itemize}
    \item We propose a novel Side-aware Meta-learning for the domain-specific tinnitus diagnosis. In the cross-dataset tinnitus classification, our model outperforms state-of-the-art methods by 2.8\%, 3.5\%, and 5.0\% on Negative-F1 score, Positive-F1 score, and accuracy, respectively.
    \item We propose a subject-specific task sampling and an auxiliary side predictor to conduct side-aware meta-learning. We also design a sliding window and down-sampling policy to align the cross-dataset signals. We implement both SMeta and conventional meta-learning on autoencoders. SMeta achieves an increase up to 1.4\%, 6.7\%, and 3.8\% on Negative-F1 score, Positive-F1 score, and accuracy respectively, demonstrating the superiority of SMeta. 
    \item Our method yields a 2.5\%/5.0\% improvement in left/right ear prediction accuracy. We conduct extensive experiments and ablation studies to show the robustness against the selection of hyper-parameters and good explainability.
\end{itemize}

\section{Related Work}
\subsection{EEG-based Tinnitus Diagnosis}

About 10\%-15\% of humans suffer from tinnitus, making tinnitus a common disorder~\cite{langguth2013tinnitus}. However, due to many possible causes, such as head injury, stress, etc., as well as different symptoms, such as hearing loss, noise trauma, etc., there is currently no universally effective clinical method for subjective tinnitus diagnosis and treatment~\cite{hall2016systematic}. Because tinnitus is often associated with changes in the brain, many researchers have proposed that the assessment of abnormal neural activity as assessed by EEG signals may aid in the diagnosis of tinnitus.

In the early stages, evaluation of EEG signals and further diagnosis were usually done by clinical specialists~\cite{7yearsfeedbackbutwork,focusonsinglebetaband}. 
Gosepath et al.~\cite{7yearsfeedbackbutwork} first attempted to use neurofeedback to assist tinnitus therapy. With the decreased activity of EEG observed, all patients claimed tinnitus relief.
The researchers then further used statistical tools to quantitatively analyze EEG data and help appraise patients' recovery condition~\cite{zscoretinnitus,scpprotocal}.
For example, Weiler et al. \cite{zscoretinnitus} proved the correlation between tinnitus and alpha, delta, and theta bands, by comparing their z-scores with control subjects.
Milner et al. \cite{scpprotocal} analyzed EEG signals quantitatively by sequentially performing high- and low-pass filtering, independent component analysis (ICA), fast Fourier transform, and mean absolute amplitudes calculation of different frequency bands.
Although they have made significant progress in diagnosing tinnitus using EEG signals, their approach suffers from inconsistent results that may lead to different diagnoses or conclusions. The reason is that their research relies heavily on the judgment of human experts, the analysis is case-specific, and there is no uniform standard.

Recent research uses machine learning and deep learning to reduce reliance on experts and mitigate the influence of personal factors in the diagnosis process~\cite{basic-tinnitus-subject-dependent,quiet-svm-tinnitus}. 
Sun et al.~\cite{basic-tinnitus-subject-dependent} segmented EEG data to enrich training sets and use Support Vector Machine (SVM) for tinnitus classification.
Li et al. \cite{quiet-svm-tinnitus} also used SVM as the classifier. They pre-processed EEG signals by first transforming the signals into the frequency domain and then performing nonlinear cosine mapping.
The machine-learning-based methods can achieve better or comparable performance than traditional diagnoses accomplished by experts, demonstrating the effectiveness of introducing machine learning into EEG analysis.

However, the testing samples of this methods~\cite{quiet-svm-tinnitus,basic-tinnitus-subject-dependent} may belong to the same subjects as training samples. The subject-dependent sampling strategy can result in similar training and testing samples and thus lead to inflated performance. To address the problem, some efforts are made to conduct subject-independent experiments, which aims to distinguish tinnitus patients from control subjects.
For example, Wang et al. \cite{power_value} adopted Fast Fourier Transform (FFT) to obtain multiple views of features from EEG signals, utilized Multi-view Intact Space Learning (MISL) to obtain latent representations, and classified samples with the Least Squares Support Vector Machine (LS-SVM).

\subsection{Cross-dataset EEG Research In Related Fields }
In addition to its application in tinnitus, EEG has also attracted much attention in other fields, where advanced deep learning methods are used~\cite{supratak2017deepsleepnet, AEXGB, liu2021task, hartmann2018eeg}.
For example, DeepSleepNet~\cite{supratak2017deepsleepnet} used convolutional neural networks (CNN) and bidirectional-long short-term-memory (LSTM) to automatically score sleep stage based on EEG signals.
Zhang et al.~\cite{AEXGB} and Hartmann et al.~\cite{hartmann2018eeg} both employed generative models, i.e., Auto-encoder and Generative adversarial networks (GANs), respectively, to synthesize EEG samples for downstream tasks, such as brain activity recognition, EEG super-sampling, or data recovery.
However, their models are optimized on large-scale datasets, limiting their deployment in more practical scenarios.

To address this issue, some cross-dataset EEG studies are proposed ~\cite{cimtay2020investigating,xu2020cross,rayatdoost2018cross}. They use the knowledge learned in other datasets to assist learning in the target dataset, naturally solving the problem of insufficient data.
However, due to the non-stationary nature of EEG data and varying distributions across subjects and datasets, the performance of directly applying deep learning models to cross-dataset problems may decrease significantly.
Cimtay et al.~\cite{cimtay2020investigating} adopted CNN to avoid manual feature extraction and find latent features, yielding impressive cross-subject and cross-dataset accuracy.
Xu et al.~\cite{xu2020cross} proposed online pre-alignment to eliminate cross-dataset variability.
Some studies~\cite{lan2018domain,lin2019constructing} further adopted the domain adaptation or transfer learning to better reuse learned knowledge in new datasets.
For example, Lan et al.~\cite{lan2018domain} employed domain adaptation to reduce the inter-subject as well as inter-datasets variance and train and test the classifier on different datasets. In contrast, Lin et al.~\cite{ lin2019constructing} integrated transfer learning and principal component analysis (PCA) to reduce the variance.

\subsection{Meta-learning Work}

In recent years, deep learning has achieved great success in various scenarios, e.g., computer vision~\cite{liu2021task,li2021attribute}, natural languages processing~\cite{gu2021domain,maulud2021state}, applications in health domains~\cite{li2020non,liu2021generalizable}. However, the advances in deep learning are based on large-scale datasets(e.g., ImageNet~\cite{deng2009imagenet}), which is not available in many scenarios of medical applications. Meta-learning~\cite{naik1992meta}, simulating humans learning how to learn, is proposed to address the problem. It aims to learn novel concepts quickly through accumulated knowledge learned from other tasks. 

A typical solution of meta-learning, called \textit{metric-based} methods, is to learn a metric or distance function to measure the similarity between new inputs and prior samples~\cite{koch2015siamese,snell2017prototypical,sung2018learning,vinyals2016matching,satorras2018few}. 
For example, Koch et al.~\cite{koch2015siamese} proposed to use Siamese Neural Networks to generate pairwise similarity between inputs and thus transfer prediction ability to data from unknown distributions. While Snell et al.~\cite{snell2017prototypical} learned a new metric space and calculated the similarity between inputs and learned prototypes of classes in the space.
Despite the advances achieved, the metric-based meta-learning fails to utilize knowledge from new tasks to update the meta-model.

Recently, some studies focus on \textit{model-based} meta-learning to learn task representations to alter internal state of models~\cite{santoro2016meta,munkhdalai2017meta,mishra2017simple,duan2016rl,wang2016learning}. The dynamic state can reflect task-specific knowledge and can thus be used to predict new inputs. 
For example, MANN~\cite{santoro2016meta} was a neural network with augmented external memory to encode and retrieve information of new data sequentially for prediction. Mishra~\cite{mishra2017simple} et al. integrated temporal convolutions and soft attention to accumulate information and pinpoint memories, respectively.
A drawback of model-based meta-learning is that it requires extra memory to remember information from previous inputs.

More related to our models are \textit{optimization-based} meta-learning~\cite{andrychowicz2016learning,ravi2016optimization,finn2017model,grant2018recasting,finn2019online}, pioneered by Andrychowicz et al. ~\cite{andrychowicz2016learning}. These models adjust the optimization process to fulfill fast gradient propagation with few training samples. Andrychowicz et al. first proposed considering the optimization procedure as a learning problem and utilizing LSTMs to replace hand-designed optimizers. Subsequently, Finn~\cite{finn2017model} et al. proposed MAML, which attracts much attention as it is simple and effective. MAML samples multiple tasks and then adopts a meta-training, meta-testing, and fine-tuning procedure to learn initial parameters. These parameters can then be fine-tuned quickly to fit new tasks.

In this paper, we adapt MAML to train our model. The reasons are three-fold: 1) MAML can learn generalizable knowledge for better cross-dataset transfer; 2) the fine-tuning process in MAML can make full use of the side information; 3) the tinnitus datasets are tiny, MAML can enrich the datasets by constructing diverse tasks.

\begin{figure*}[h]
\centering
\includegraphics[width=0.88\textwidth]{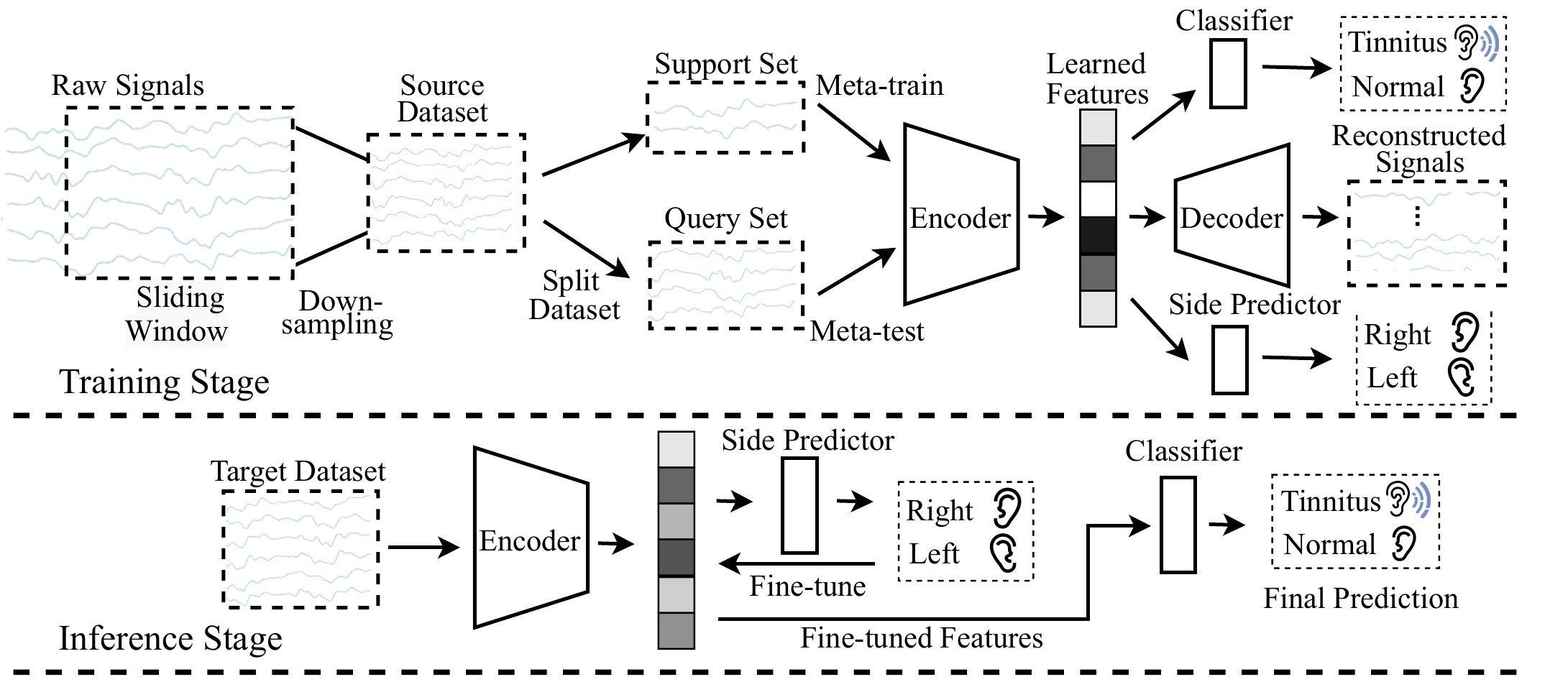}
\caption{Model overview. The upper part shows the training stage of SMeta, and the lower part draws the data flow of the inference stage. In our experiments, signals from the target dataset have fewer time points than those from source datasets. Therefore, the training stage first uses a sliding window and down-sampling to align the data format of the source dataset to the target dataset. Then, the down-sampled source dataset will be split into two sub-sets for meta-training and meta-testing, respectively. Both meta-training and meta-testing not only learn to reconstruct raw signals and classify signals but also learn tinnitus-specific information, i.e., the side information of left/right ears. The inference stage uses the trained encoder from the training stage and further fine-tunes the encoder based on side predictor fitting left/right ears. Then, we take the fine-tuned features for the final prediction.}
\label{Overview}
\end{figure*}

\section{Methodology}

\subsection{Problem Definition}
Given a source dataset $S$ and a target dataset $T$ with different data formats, $x_{i}\in S\cup T$ denotes an EEG signal. Specifically, $x^{s}_{i}$ and $x^{g}_{i}$ are signals sampled from the source and the target dataset, respectively. Each signal consists of $n$ time points $x_{i}=\{o_{j}:j\in[1,n]\}$, has a corresponding label $y^{t}_{i}$ for tinnitus diagnosis, and has a side label $y^{d}_{i}$ to record the left/right ear where the signal is collected. $y^{t}_{i}=1$ denotes positive subject (i.e., tinnitus patient) and $y^{t}_{i}=0$ denotes negative subject (i.e., control subject). $y^{d}_{i}=1$ denotes right ear and $y^{d}_{i}=0$ denotes left ear. Our goal is to learn a unified encoder $f_{en}(x_{i})\rightarrow e_{i}$ and a unified classifier $f_{c}(e_{i})\rightarrow y^{t}_{i}$ that are effective for both source and target datasets, where $e_{i}$ denotes the extracted feature of $x_{i}$. To learn tinnitus-specific information of left/right ear, we design a side predictor $f_{p}(e_{i})\rightarrow y^{d}_{i}$ to predict the side information. To regularize the learned feature, we use a decoder $f_{de}(e_{i})\rightarrow x_{i}$ to restrain $e_{i}$ from over-fitting classification information. The model overview is shown in Fig.~\ref{Overview}.

\subsection{Dataset Pre-processing}

Different data formats may cause the problems that signals from source dataset $S$ and target dataset $T$ sample different numbers of sampled time points (i.e., sampling frequency), use different filter policies in the same time duration or have divergent strength range due to the different experimental equipment. Suppose that the signals that have the same time duration in source and target datasets are with $n_{s}$ and $n_{g}$ time points ($n_{s}\geq n_{g}$), respectively. We can use sliding window to cut source signals $x^{s}_{i}$ into slices $\{x^{o}_{1},x^{o}_{2},...\}$ that have the same time duration as $x^{g}_{i}$. Then, we can use down-sampling to align the length of time points in the source dataset to those in the target dataset as follows:

\begin{gather}\label{slide_down_sample}
\bar{o}^{s}_{j}=\left\{\begin{matrix}
\frac{1}{l}\sum_{k\in [jl-l+1,jl]}o^{s}_{k} & j\in[1,m]\\ 
\frac{1}{l+1}\sum_{k\in [jl+j-m-l,jl+j-m+1]}o^{s}_{k} & j\in[m+1,n_{g}]
\end{matrix}\right.\\
s.t. \quad l = \left \lfloor n_{s}/n_{g} \right \rfloor \quad m = n_{s}-l*n_{g}
\end{gather}
where $o^{s}_{k}$ denotes $k^{th}$ original time point in the sliced source signal $x^{o}_{i}$; $\bar{o}^{s}_{j}$ denotes the new averaged time point for the sliced source signal; the window size is $n_{s}$; $\left \lfloor n_{s}/n_{g} \right \rfloor$ denotes the floor of $n_{s}/n_{g}$.

After sliding and down-sampling, $\bar{o}^{s}_{j}$ will be the new time points in source signals, where the front $m$ values in new signals of the source dataset will be the mean values of $l$ time points; the last $(n_{g}-m)$ values will be the mean values of ($l+1$) time points.

Then, we use min-max normalization to separately limit the signal strengths to the same value range as follows:

\begin{gather}\label{min_max_norm}
    \hat{o}^{s}_{j}=\frac{\bar{o}^{s}_{j}-\min \bar{o}^{s}_{j}}{\max \bar{o}^{s}_{j} - \min \bar{o}^{s}_{j}}\\
    \hat{o}^{g}_{j}=\frac{o^{g}_{j}-\min o^{g}_{j}}{\max o^{g}_{j} - \min o^{g}_{j}}
\end{gather}
where $o^{g}_{j}$ denotes the time points of the signal from the target dataset (i.e., $x^{g}_{i}\in T$; $\hat{o}^{s}_{j}$ and $\hat{o}^{g}_{j}$ represent the normalized time points in the source signal and the target signal, respectively.

We let $\hat{x}^{s}_{i}$ and $\hat{x}^{g}_{i}$ be the new source and target signals constituted by $\bar{o}^{s}_{j}$ and $o^{g}_{j}$, respectively. $\hat{x}^{s}_{i}$ will have the same length (i.e., $n_{g}$) as the signal $\hat{x}^{g}_{i}$ for the target dataset. Note that when $n_{s}<n_{g}$, we can use the same operations to align the target dataset to the source dataset by sliding window and down-sampling.

\begin{figure}[h]
\centering 
\includegraphics[width=0.48\textwidth]{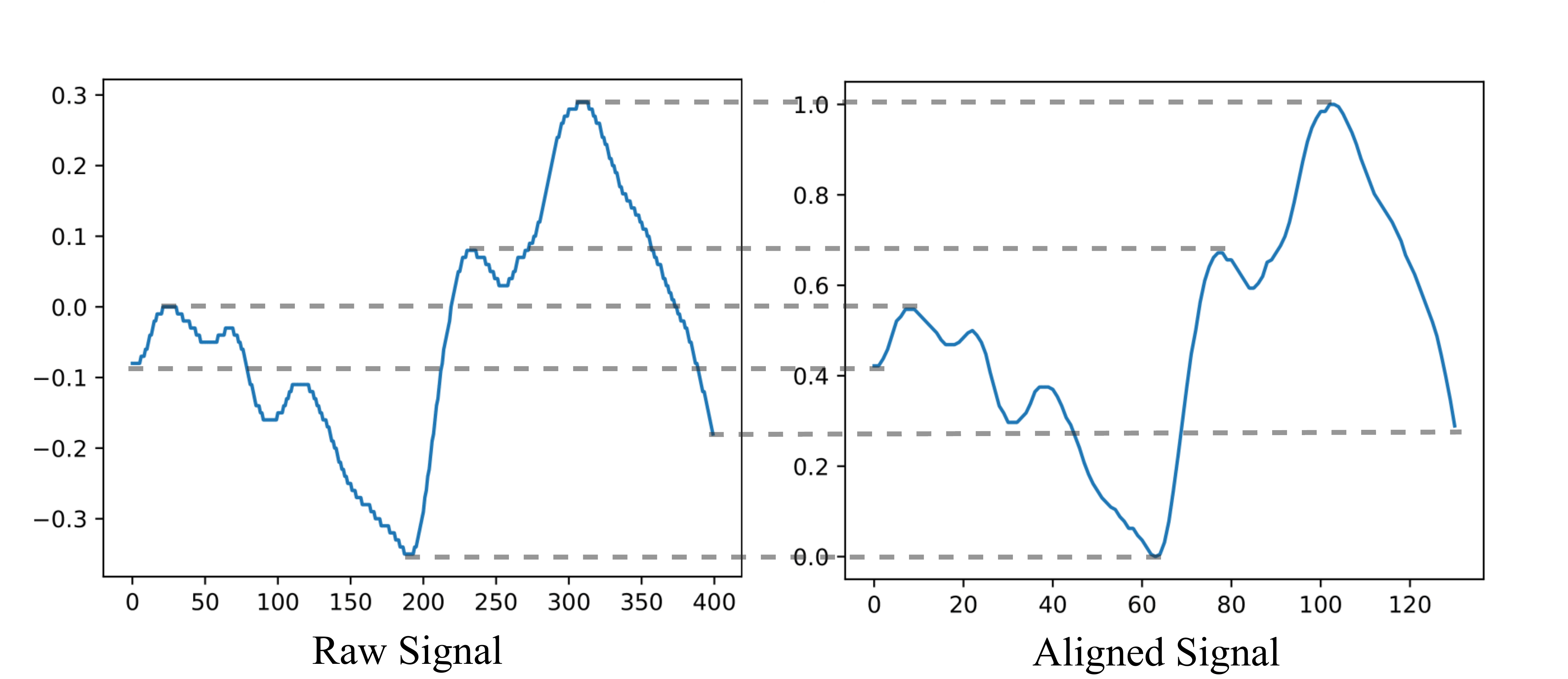}
    \caption{Illustration of the same signal trend in (left) the raw signal and (right) the aligned signal, i.e., the normalized and down-sampled sliding window. x-axis denotes the time points and y-axis denotes the signal strength.}\label{raw_and_slided}
\end{figure}

As shown in Figs.~\ref{raw_and_slided}, we take the 1) raw signal and 2) the normalized down-sampled signal as an example. From the parallel dashes, we can observe that the down-sampling and min-max normalization will squeeze the signal strength to a unified value range $(0,1)$ but not change the original signal trend or pattern.

\subsection{Side-aware Meta-autoencoder}
We let $\hat{S}=\{\hat{x}^{s}_{1},\hat{x}^{s}_{2},...\}$ be the normalized and down-sampled source dataset. Following the meta-learning setting~\cite{finn2017model} and the subject-independent setting~\cite{liu2021generalizable}, we propose a subject-specific task sampling strategy for tinnitus analysis. Regarding the significant subject variance, we view each subject as a task and denote a subject $k$ in the source dataset by $\tau_{k} \in \hat{S}$. Then, we conduct episode-wise meta-learning.

In each episode, we sample $b$ tasks (i.e., independent subjects) from $\hat{S}$, where $b$ denotes batch size. For each subject $\tau_{k}$, we randomly split the signals into two disjoint sets: support set $\tau^{spt}_{k}$ and query set $\tau^{qry}_{k}$ for meta-training and meta-learning, respectively. Note that $\tau^{spt}_{k}\cap \tau^{qry}_{k}=\emptyset$. In the meta-learning phase, we design three loss functions to learn different aspects of signal information: classification loss $L_{cls}$, reconstruction loss $L_{rec}$, and ear side prediction loss $L_{ear}$.

Classification loss $L_{cls}$ aims to supervise the learned features to carry the information to accurately classify tinnitus patients and control subjects:
\begin{equation}\label{cls_loss}
    L_{cls}=CrossEntropy(f_{c}(f_{en}(\hat{x}^{s}_{i})),y^{s}_{i})
\end{equation}
where $CrossEntropy$ denotes cross entropy loss; $y^{s}_{i}$ is the corresponding ground-truth label of $\hat{x}^{s}_{i}$.

We design reconstruction loss $L_{rec}$ to enable the learned features to carry the waveform information to reconstruct the original signals, which can prevent the encoder from over-fitting class information:
\begin{equation}\label{rec_loss}
    L_{rec}=MSE(f_{de}(f_{en}(\hat{x}^{s}_{i})),\hat{x}^{s}_{i})
\end{equation}
where $MSE$ represents mean square error.

Ear side prediction loss $L_{ear}$ tends to optimize the learned features to be aware of the ear side information among signals, which may assist in classifying subjects:
\begin{equation}\label{sde_loss}
    L_{ear}=CrossEntropy(f_{p}(f_{en}(\hat{x}^{s}_{i})),y^{d}_{i})
\end{equation}
where $y^{d}_{i}$ is the corresponding ground-truth ear side of $\hat{x}^{s}_{i}$.

Then, we can summarize the loss function for optimizing our Side-aware Meta-AutoEncoder (SMeta-AE) as follows:
\begin{equation}\label{smeta_ae_loss}
    L_{SMeta\hbox{-}AE}=L_{cls}+L_{rec}+L_{ear}
\end{equation}

Based on the SMeta-AE loss, we first meta-train the model based on the support set, which virtually optimizes the model to learn a basic model. Then, we meta-test the model based on the query set, which summarizes the overall gradients of tasks to truly optimize the model and thus obtains generalized model parameters. The meta-learning can be summarized as follows:
\begin{equation}\label{smeta-ae-meta-train}
    \theta_{AE}^{'}(\tau^{spt}) \leftarrow \theta_{AE}-\alpha\sum_{k}\bigtriangledown_{\theta_{AE}}\mathcal{L}^{\tau^{spt}_{k}}_{SMeta\hbox{-}AE}(\theta_{AE})
\end{equation}
\begin{equation}\label{smeta-ae-meta-test}
    \theta_{AE}^{*}\leftarrow\theta_{AE}-\beta\sum_{k}\bigtriangledown_{\theta_{AE}}\mathcal{L}^{\tau^{qry}_{k}}_{SMeta\hbox{-}AE}(\theta_{AE}^{'}(\tau^{spt}))
\end{equation}
where $\theta_{AE}$ are model parameters of SMeta-AE; $\theta_{AE}^{'}(\tau^{spt})$ represents the virtual parameters using support sets to meta-train the model; $\theta_{AE}^{*}$ are the updated parameters based on the query sets; $\bigtriangledown_{\theta_{AE}}\mathcal{L}^{\tau^{spt}_{k}}_{SMeta\hbox{-}AE}(\theta_{AE})$ represents the gradient calculation of SMeta-AE for loss $L_{SMeta\hbox{-}AE}$ using $\theta_{AE}$ as original model parameters and $\tau^{spt}_{k}$ as the training samples; $\alpha$ is the meta-learning rate for meta-training; $\beta$ is the learning rates for meta-testing, respectively. 


\subsection{Siamese Autoencoder Extension}
In this section, we introduce a Siamese autoencoder extension, named Side-aware Meta-Siamese-AutoEncoder (SMeta-SAE), to our basic version of autoencoder following the work of Liu et al.~\cite{liu2021generalizable}. Additionally, we design a subject predictor $f_{sub}(e_{i},e_{j})\rightarrow (0,1)$ to predict which subject the signals belong to. Note that we simplify the contrastive learning about subject prediction to two subjects, which is sufficient for meta-learning. To conduct pairwise training for episode-wise meta-learning, we sample two subjects each time, and fuse their support and query sets in a half-to-half way: keeping half of the sets without fusion as samples that belong to the same subjects and fusing the last half of the sets of two subjects as samples that belong to the different subjects. Then, we design the loss function for SMeta-SAE based on an adversarial training loss as follows:
\begin{equation}
\begin{gathered}
L_{SMeta\hbox{-}SAE}=L_{SMeta\hbox{-}AE}+L_{adv}+L_{sub}\\
=L_{cls}+L_{rec}+L_{ear}+\gamma_{i,j}*MSE(e_{i},e_{j})\\
    +CrossEntropy(f_{sub}(e_{i},e_{j}),u_{i,j}) 
\end{gathered}
\end{equation}
where $L_{SMeta\hbox{-}AE}$ denotes the sum of the common loss functions $L_{cls}$, $L_{rec}$, and $L_{ear}$; $\gamma_{i,j}$ is a sign function that equals 1 when $i=j$ otherwise -1; $u_{i,j}$ is the contrastive label which equals 1 when $i=j$ otherwise 0.

$L_{adv}$ is an adversarial loss. It aligns the learned features when they are form the same subject; otherwise it optimizes the learned features to be divergent. $L_{sub}$ is a contrastive label that enhances the model learning abilities by comparing signals from different subjects.

Then, we can easily infer the meta-learning loss for SMeta-SAE as follows:
\begin{equation}\label{smeta-sae-meta-train}
    \theta_{SAE}^{'}(\tau^{spt}) \leftarrow \theta_{SAE}-\alpha\sum_{k}\bigtriangledown_{\theta_{SAE}}\mathcal{L}^{(\tau^{spt}_{k,1},\tau^{spt}_{k,2})}_{SMeta\hbox{-}SAE}(\theta_{SAE})
\end{equation}
\begin{equation}\label{smeta-sae-meta-test}
    \theta_{SAE}^{*}\leftarrow\theta_{SAE}-\beta\sum_{k}\bigtriangledown_{\theta_{SAE}}\mathcal{L}^{(\tau^{qry}_{k,1},\tau^{qry}_{k,2})}_{SMeta\hbox{-}SAE}(\theta_{SAE}^{'}(\tau^{spt}))
\end{equation}
where $\theta_{SAE}$ are model parameters of SMeta-SAE. 


\subsection{Training and Inference Strategy for Meta-learning}
In the training stage, we first train the AE and SAE to get a basic model. Then, we take the trained model as initial parameters to conduct conventional meta-learning and side-aware meta-learning. We do not use random parameters as an initial model for meta-learning because meta-learning will weaken the fitting ability of models. The random initialization may cause the model under-fitting.

In the inference stage, we lack observed samples for the conventional meta-learning, so we directly use the trained model to conduct inference. For side-aware meta-learning, as shown in the lower part of Fig.~\ref{Overview}, we use side information from the target dataset to fine-tune the model parameters for each subject as follows:
\begin{equation}
    \theta_{k} \xleftarrow{fine-tune} \theta-\beta\bigtriangledown_{\theta}\mathcal{L}_{ear}^{\tau^{test}_{k}\in \hat{T}}
\end{equation}
where $\theta_{k}$ denotes the fine-tuned model parameters of AE or SAE for subject $k$; $\beta$ is the learning rate; $\hat{T}=\{\hat{x}^{g}_{1},\hat{x}^{g}_{2},...\}$.

\section{Experiment}
\begin{table*}
\centering
\caption{Best performance of cross-dataset tinnitus diagnosis on both ears (standard deviation).}
\label{Main_exp_both}
\small
\begin{tabular}{c|ccccccc} 
\toprule
\multirow{2}{*}{Model} & \multicolumn{7}{c}{Both Sides}                                                                          \\ 
\cmidrule{2-8}
                                                                      & NPV          & TNR          & N-F1         & PPV          & TPR          & P-F1         & Acc           \\ 
\midrule
XGBoost                                                               & 0.467(0.000) & 0.525(0.000) & 0.494(0.000) & 0.457(0.000) & 0.400(0.000) & 0.427(0.000) & 0.463(0.000)  \\
Nu-SVC                                                                & 0.467(0.000) & 0.350(0.000) & 0.400(0.000) & 0.480(0.000) & 0.600(0.000) & 0.533(0.000) & 0.475(0.000)  \\
nCSP                                                                  & 0.568(0.055) & 0.625(0.078) & 0.595(0.058) & 0.583(0.054) & 0.525(0.085) & 0.553(0.063) & 0.575(0.053)  \\
DeepNet                                                               & 0.714(0.243) & 0.125(0.042) & 0.213(0.070) & 0.521(0.010) & 0.950(0.019) & 0.673(0.009) & 0.538(0.018)  \\
ShalowNet                                                             & 0.581(0.044) & 0.625(0.147) & 0.602(0.112) & 0.595(0.033) & 0.550(0.110) & 0.571(0.040) & 0.588(0.035)  \\
AE-XGB                                                                & 0.696(0.071) & 0.400(0.170) & 0.508(0.158) & 0.579(0.033) & 0.825(0.126) & 0.680(0.041) & 0.613(0.041)  \\
EEGNET                                                                & 0.605(0.081) & 0.650(0.230) & 0.627(0.181) & 0.622(0.048) & 0.575(0.159) & 0.597(0.055) & 0.613(0.049)  \\
AE                                                                    & 0.660(0.047) & 0.775(0.077) & 0.713(0.032) & 0.727(0.104) & 0.600(0.154) & 0.658(0.143) & 0.688(0.056)  \\
SAE                                                                   & 0.703(0.029) & 0.650(0.063) & 0.675(0.018) & 0.674(0.027) & 0.725(0.076) & 0.699(0.034) & 0.688(0.013)  \\
\midrule
Meta-AE                                                               & 0.667(0.013) & 0.800(0.036) & 0.727(0.012) & 0.750(0.020) & 0.600(0.036) & 0.667(0.014) & 0.700(0.007)  \\
Meta-SAE                                                              & 0.681(0.018) & 0.800(0.045) & 0.736(0.014) & 0.758(0.023) & 0.625(0.048) & 0.685(0.020) & 0.713(0.010)  \\
SMeta-AE                                                         & 0.714(0.020) & 0.750(0.025) & 0.732(0.010) & 0.737(0.012) & 0.700(0.040) & 0.718(0.022) & 0.725(0.013)  \\
SMeta-SAE                                                          & 0.732(0.019) & 0.750(0.061) & 0.741(0.022) & 0.744(0.040) & 0.725(0.053) & 0.734(0.020) & 0.738(0.013)  \\
\bottomrule
\end{tabular}
\end{table*}
\subsection{Experiment Setting}\label{exp_set}
In the experiments, we examine the efficacy of our method using two tinnitus dataset~\cite{schaette2011tinnitus,guest2017tinnitus} through Auditory brainstem responses (ABRs) through EEG. ABRs are evoked potentials recorded with EEG sensors, which originate from the early stages of the auditory pathway. The source dataset was collected by Schaette and McAlpine~\cite{schaette2011tinnitus}, which used presentation levels of 90 and 100 dB peSPL with an 11 click/s rate. The target dataset was collected by Guest et al.~\cite{guest2017tinnitus} in response to 102 dB peSPL clicks presented with a rate of 7 clicks/s. The source dataset was approved by the University College London (UCL) ethics committee with ethics ID number 2039/002. The declaration is the Helsinki declaration. The target dataset was approved by the National Research Ethics Service Greater Manchester West Ethics Committee with REC reference 15/NW/0133 and IRAS project ID 168221. The source dataset~\cite{schaette2011tinnitus} contains 408 signals of a duration of 10 ms from 38 subjects, where each subject only has data of one side (left or right ear). The target dataset~\cite{guest2017tinnitus} has 80 signals of a duration of 8 ms from 40 subjects, where each subject in the target dataset has records for both left and right ears, respectively. In other words, we use 408 signals of the source dataset as the training data, and 80 signals of the target dataset as the testing data. Since we use different datasets for training and testing, K-fold cross-validation is not applicable for our experiments. The source dataset only consists of the female subjects and the target dataset contains both male and female subjects. The averaged time point numbers are 500 and 131 for the source and target datasets, respectively. The mean ages of patients/subjects are 36.3/33.2 years and 25.7/25.5 years in the source dataset and target dataset, respectively. Both datasets are measured by a Medelec Synergy T-EP system (Oxford Instruments Medical) by placing Disposable electrodes (Nicolet Biomedical) on the high forehead and the ipsilateral and contralateral mastoids, where the electrode impedances are two k$\Omega$. 
Therefore, $n_{s}$=400 and $n_{g}$=131 to make slices of signals in the same time duration, i.e., 8 ms.

For both SMeta-AE and SMeta-SAE, we use a 1-way 2-shot learning scheme for meta-learning following the work of Finn et al.~\cite{finn2017model}. We use this setting because we want to have a subject-specific fine-tuned model for each subject to ease the negative influence of subject variance on classification, and the target dataset only has two samples for each subject, i.e., 2 samples to fit in the inference stage. The window interval is 20 when sampling signals in the source dataset. We set $\alpha$=1e-3, $\beta$=1e-3, query size as 8, and batch size $b$=16 for SMeta-AE and $b$=5 for SMeta-SAE. The epoch maximum is 200 for SMeta-AE during training. To allow more training times for the Siamese autoencoder which is more complex than a conventional autoencoder, we set the epoch maximum as 1000 for SMeta-SAE.

We compare our method with 9 competitive machine learning models: (\textbf{a}) XGBoost~\cite{chen2016xgboost}, (\textbf{b}) nu-SVC~\cite{v-svm}, (\textbf{c}) nCSP~\cite{ncsp}, (\textbf{d}) DeepNet \cite{DeepConvNet}, (\textbf{e}) ShallowNet \cite{DeepConvNet} (\textbf{f}) AEXGB \cite{AEXGB}, (\textbf{g}) EEGNet \cite{eegnet}, (\textbf{h}) AE~\cite{sunilkumar2021bio}, and (\textbf{i}) SAE~\cite{liu2021generalizable}. We view tinnitus patients as positive samples and control subjects as negative samples. We use criteria, Negative Predictive Value (NPV), True Negative Rate (TNR), Positive Predictive Value (PPV), True Positive Rate (TPR), F1-score for negative samples (N-F1), and F1-score for positive samples (P-F1), to measure the model performance among positive and negative samples. The calculation is as follows:
\begin{gather}
    NPV = TN / (FN + TN)\\
    TNR = TN / (FP + TN)\\
    PPV = TP / (TP + FP)\\
    TPR = TP / (TP + FN)\\
    N\hbox{-}F1: 2*NPV*TNR/(NPV+TNR)\\
    P\hbox{-}F1: 2*PPV*TPR/(PPV+TPR)
\end{gather}
where TP, TN, FP, FN represent the numbers of true positive, true negative, false positive, and false negative prediction, respectively.

\begin{figure*}[h]
\centering 
\begin{subfigure}{0.32\textwidth}
  \includegraphics[width=\textwidth]{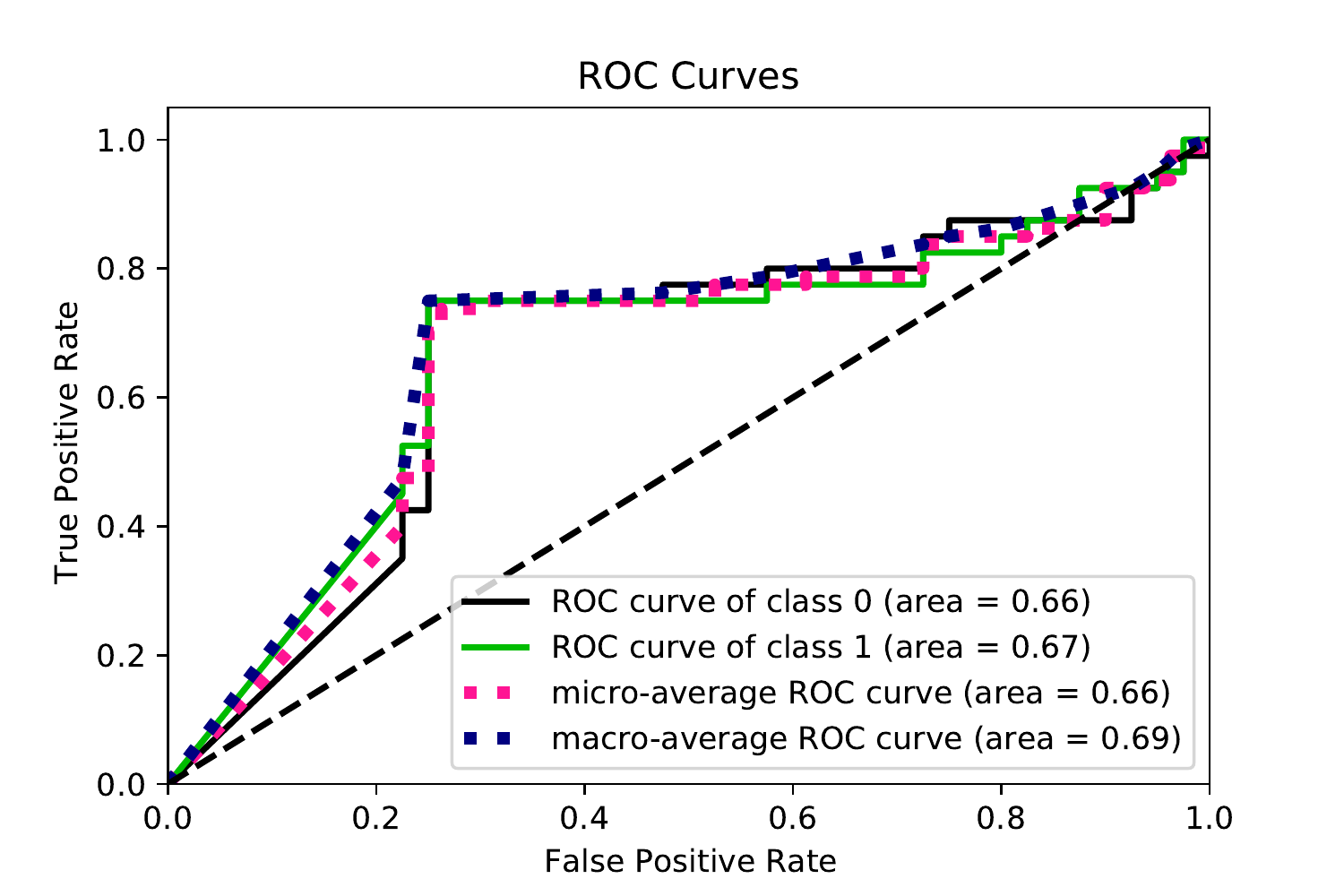}
    \centering
  \caption{ROC curve of both sides.}
\end{subfigure}\hfil 
\begin{subfigure}{0.32\textwidth}
  \includegraphics[width=\textwidth]{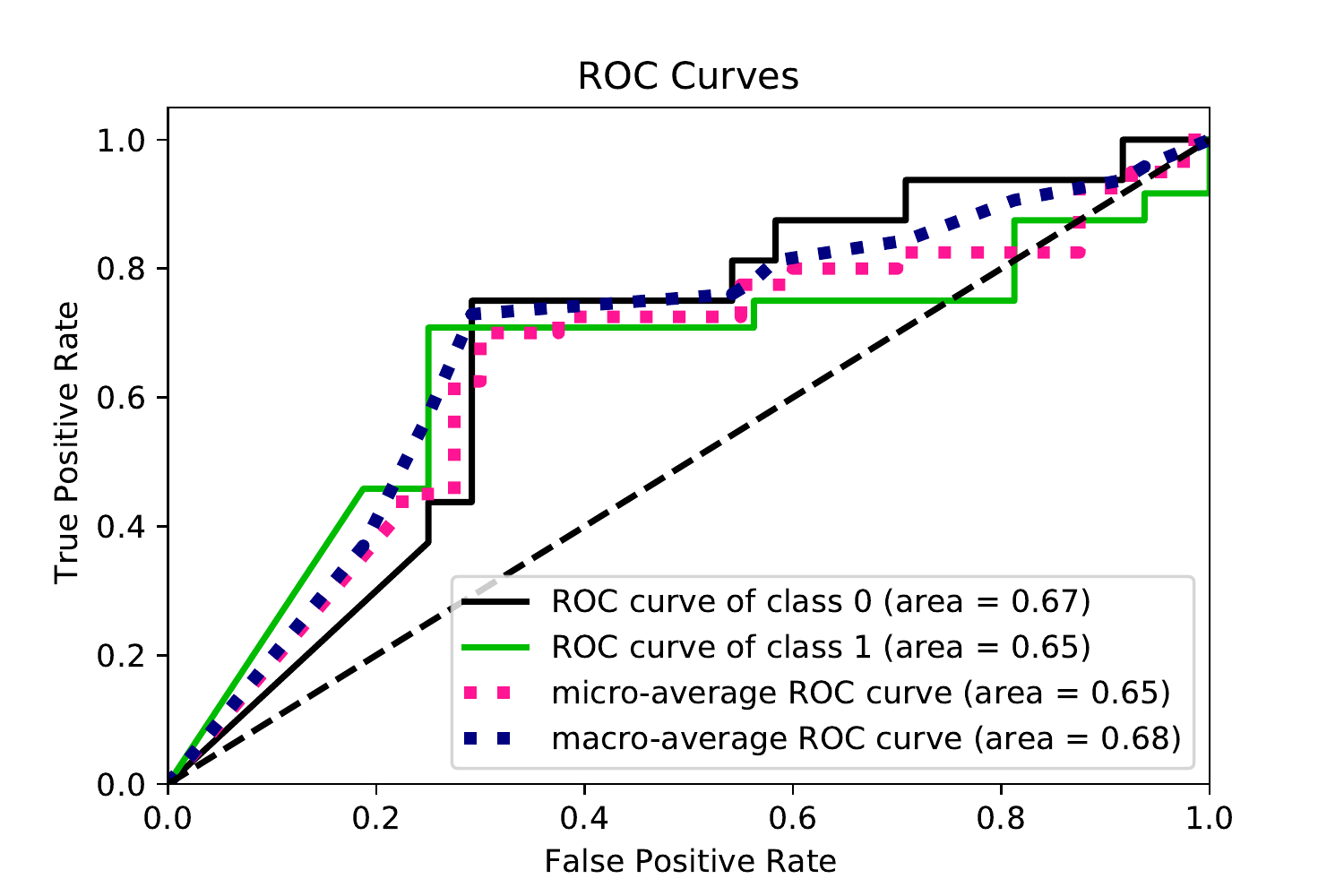}
    \centering
  \caption{ROC curve of left side.}
\end{subfigure}\hfil 
\begin{subfigure}{0.32\textwidth}
  \includegraphics[width=\textwidth]{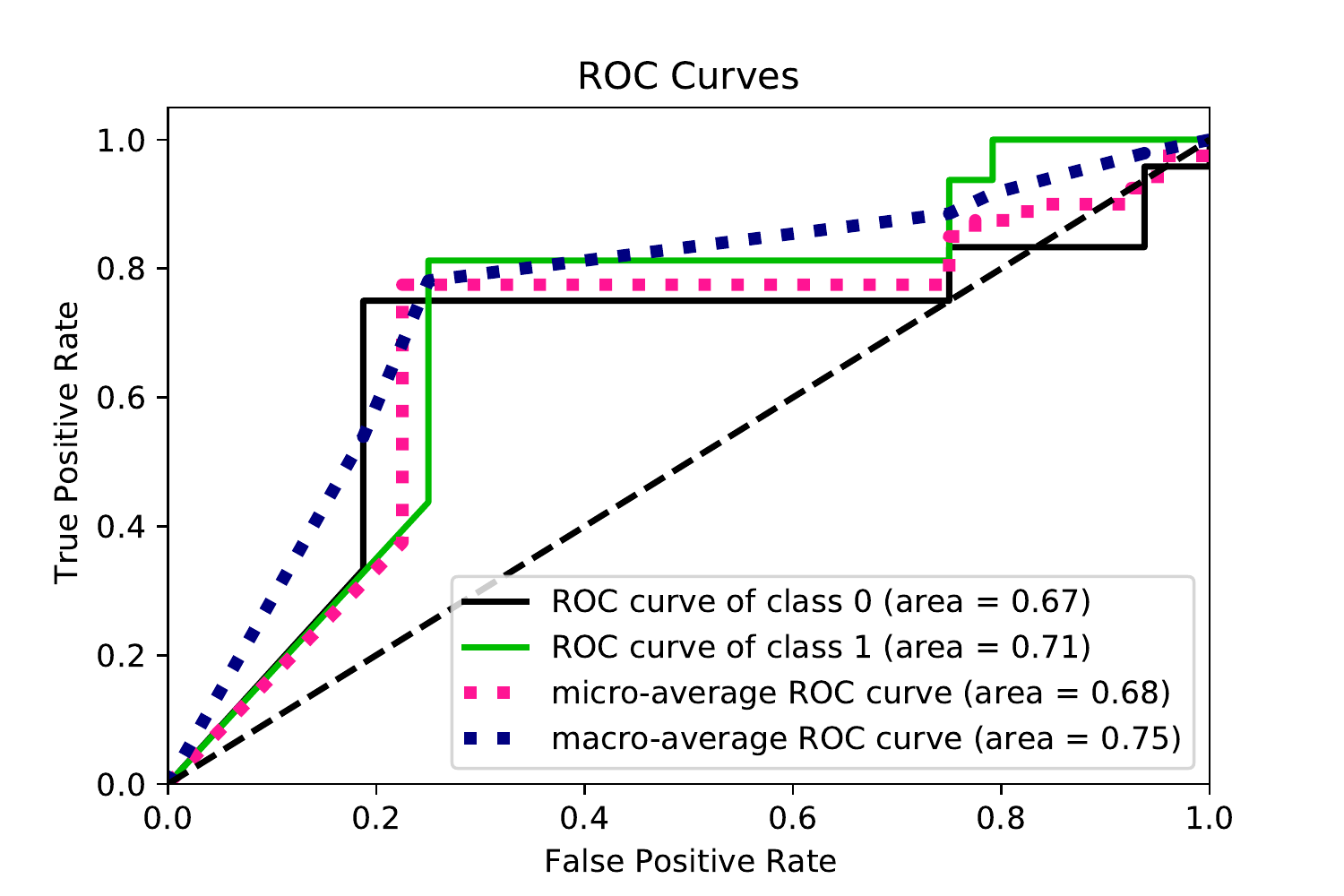}
    \centering
  \caption{ROC curve of right side.}
\end{subfigure}\hfil 
    \caption{ROC curves of SMeta-SAE. Class 0 and class 1 denote control subject and tinnitus patient, respectively.}\label{ROC_curve}
\end{figure*}

\subsection{Cross-dataset Tinnitus Diagnosis}
We run each model 50 times and use the best performance as the final result. The best performance and standard deviation are shown in Table~\ref{Main_exp_both}. In the compared methods, AE and SAE achieve the highest accuracy. Meanwhile, AE obtains the highest scores in TNR, N-F1, and PPV, which means that AE has a better ability in classifying negative samples than SAE. SAE has the best scores in TPR and P-F1, which shows that SAE is effective in classifying positive samples. DeepNet gets the highest score in NPV. Overall, deep-learning-based methods have the best performance in machine learning methods. We show both conventional meta-learning and side-aware meta-learning on AE and SAE. Compared to these methods, we can observe that both meta-learning and side-aware meta-learning can effectively improve the model performance. Specifically, SMeta-based method can improve up to 1.8\%, 2.8\%, 1.7\%, 3.5\%, 5.0\% in NPV, N-F1, PPV, P-F1, and ACC, respectively. The meta-based method can improve up to 2.5\%, 2.3\%, 3.1\%, 2.5\%, in TNR, N-F1, PPV, and Acc, respectively. Meta-learning mainly improves the ability to classify negative samples, while SMeta can improve the ability to classify both positive and negative samples. However, SMeta methods tend to be less stable than Meta methods, which may be caused by the over-fitting towards side information in the inference stage.

\begin{table*}
\centering
\caption{Best performance of  cross-dataset tinnitus diagnosis on the left ear (standard deviation).}
\label{Main_exp_left}
\small
\begin{tabular}{c|ccccccc} 
\toprule
\multirow{2}{*}{\begin{tabular}[c]{@{}c@{}}\\Model\end{tabular}} & \multicolumn{7}{c}{Left Side}                                                                          \\ 
\cmidrule{2-8}
                                                                 & NPV          & TNR          & N-F1         & PPV          & TPR          & P-F1         & Acc           \\ 
\midrule
XGBoost                                                          & 0.455(0.000) & 0.500(0.000) & 0.476(0.000) & 0.444(0.000) & 0.400(0.000) & 0.421(0.000) & 0.450(0.000)  \\
Nu-SVC                                                           & 0.467(0.000) & 0.350(0.000) & 0.400(0.000) & 0.480(0.000) & 0.600(0.000) & 0.533(0.000) & 0.475(0.000)  \\
nCSP                                                             & 0.591(0.088) & 0.650(0.109) & 0.619(0.090) & 0.611(0.078) & 0.550(0.114) & 0.579(0.091) & 0.600(0.081)  \\
DeepNet                                                          & 0.500(0.245) & 0.100(0.053) & 0.167(0.085) & 0.500(0.016) & 0.900(0.026) & 0.643(0.017) & 0.575(0.029)  \\
ShalowNet                                                        & 0.524(0.063) & 0.550(0.136) & 0.537(0.110) & 0.526(0.036) & 0.500(0.122) & 0.513(0.055) & 0.525(0.044)  \\
AE-XGB                                                           & 0.615(0.143) & 0.400(0.179) & 0.485(0.169) & 0.556(0.049) & 0.750(0.146) & 0.638(0.065) & 0.575(0.060)  \\
EEGNET                                                           & 0.600(0.115) & 0.600(0.240) & 0.600(0.196) & 0.600(0.054) & 0.600(0.169) & 0.600(0.068) & 0.600(0.062)  \\
AE                                                               & 0.652(0.053) & 0.750(0.078) & 0.698(0.043) & 0.706(0.157) & 0.600(0.159) & 0.649(0.161) & 0.675(0.069)  \\
SAE                                                              & 0.750(0.038) & 0.600(0.077) & 0.667(0.035) & 0.667(0.036) & 0.800(0.073) & 0.727(0.030) & 0.700(0.023)  \\ 
\midrule
Meta-AE                                                          & 0.682(0.018) & 0.750(0.034) & 0.714(0.016) & 0.722(0.019) & 0.650(0.037) & 0.684(0.020) & 0.700(0.015)  \\
Meta-SAE                                                         & 0.682(0.021) & 0.750(0.038) & 0.714(0.018) & 0.722(0.022) & 0.650(0.044) & 0.684(0.024) & 0.700(0.017)  \\
SMeta-AE                                                    & 0.714(0.030) & 0.750(0.029) & 0.732(0.023) & 0.737(0.023) & 0.700(0.047) & 0.718(0.032) & 0.725(0.026)  \\
SMeta-SAE                                                     & 0.714(0.024) & 0.750(0.056) & 0.732(0.024) & 0.737(0.036) & 0.700(0.051) & 0.718(0.023) & 0.725(0.019)  \\
\bottomrule
\end{tabular}
\end{table*}

\begin{table*}
\centering
\caption{Best performance of cross-dataset tinnitus diagnosis on the right ear (standard deviation).}
\label{Main_exp_right}
\small
\begin{tabular}{c|ccccccc} 
\toprule
\multirow{2}{*}{\begin{tabular}[c]{@{}c@{}}\\Model\end{tabular}} & \multicolumn{7}{c}{Right Side}                                                                          \\ 
\cmidrule{2-8}
                                                                 & NPV          & TNR          & N-F1         & PPV          & TPR          & P-F1         & Acc           \\ 
\midrule
XGBoost                                                          & 0.478(0.000) & 0.550(0.000) & 0.512(0.000) & 0.471(0.000) & 0.400(0.000) & 0.432(0.000) & 0.475(0.000)  \\
Nu-SVC                                                           & 0.467(0.000) & 0.350(0.000) & 0.400(0.000) & 0.480(0.000) & 0.600(0.000) & 0.533(0.000) & 0.475(0.000)  \\
nCSP                                                             & 0.545(0.068) & 0.600(0.099) & 0.571(0.077) & 0.556(0.062) & 0.500(0.109) & 0.526(0.078) & 0.550(0.063)  \\
DeepNet                                                          & 1.000(0.416) & 0.150(0.047) & 0.261(0.082) & 0.541(0.011) & 1.000(0.028) & 0.702(0.012) & 0.500(0.021)  \\
ShalowNet                                                        & 0.636(0.049) & 0.700(0.166) & 0.667(0.125) & 0.667(0.045) & 0.600(0.104) & 0.632(0.033) & 0.650(0.043)  \\
AE-XGB                                                           & 0.800(0.115) & 0.400(0.174) & 0.533(0.164) & 0.600(0.039) & 0.900(0.127) & 0.720(0.044) & 0.650(0.051)  \\
EEGNET                                                           & 0.609(0.060) & 0.700(0.226) & 0.651(0.170) & 0.647(0.054) & 0.550(0.155) & 0.595(0.050) & 0.625(0.046)  \\
AE                                                               & 0.667(0.049) & 0.800(0.098) & 0.727(0.037) & 0.750(0.110) & 0.600(0.160) & 0.667(0.140) & 0.700(0.055)  \\
SAE                                                              & 0.667(0.038) & 0.700(0.063) & 0.683(0.027) & 0.684(0.037) & 0.650(0.090) & 0.667(0.054) & 0.675(0.031)  \\ 
\midrule
Meta-AE                                                          & 0.654(0.019) & 0.850(0.048) & 0.739(0.019) & 0.786(0.034) & 0.550(0.046) & 0.647(0.022) & 0.700(0.016)  \\
Meta-SAE                                                         & 0.680(0.021) & 0.850(0.060) & 0.756(0.020) & 0.800(0.037) & 0.600(0.060) & 0.686(0.024) & 0.725(0.012)  \\
SMeta-AE                                                    & 0.714(0.025) & 0.750(0.037) & 0.732(0.017) & 0.737(0.026) & 0.700(0.050) & 0.718(0.028) & 0.725(0.019)  \\
SMeta-SAE                                                     & 0.750(0.030) & 0.750(0.073) & 0.750(0.030) & 0.750(0.053) & 0.750(0.070) & 0.750(0.035) & 0.750(0.025)  \\
\bottomrule
\end{tabular}
\end{table*}

We further show the detailed prediction of left and right ear in Table~\ref{Main_exp_left} and Table~\ref{Main_exp_right}, respectively. We can find that SMeta-based methods still achieve the best performance on both left and right ears in Acc. SMeta-SAE improves the accuracy up to 2.5\% and 5.0\% compared to the best comparison in left and right ear, respectively. In the aspect of balanced classification ability, SMeta-based methods are suitable for the negative subject in both left and right ears. It can also well handle tinnitus patients in the right ear. While meta-learning may impair the model performance of TPR in the left ear, both Meta- and SMeta-methods can make the model classify tinnitus patients with higher precision, i.e., PPV.

In conclusion, side-aware meta-learning can effectively improve the ability of the model to classify tinnitus patients and control subjects of different gender and ages. The side information of ears can boost the model training and fitting process to further enhance the meta-learning. It is feasible to use side-aware meta-learning to develop a reliable classification tool for ubiquitous patients with diverse physical conditions, e.g., age and gender, based on limited datasets. 

\subsection{Ablation Study}

\subsubsection{ROC curves} As shown in Figs.~\ref{ROC_curve} (a-c), we show the Receiver Operating Characteristic (ROC) curves of SMeta-SAE, which can achieve a high Area Under Curve (AUC) score for ROC in three conditions. When classifying both left and right sides at the same time, the areas are similar. SMeta-SAE obtains the highest AUC score in precisely classifying the right ear of tinnitus patients. Overall, SMeta-SAE shows a better AUC in the right ear than the left ear.

\begin{table*}[h]
\centering
\caption{Parameter study of SMeta-AE on batch size $b$.}
    \label{Ablation_batch_size}
    \resizebox{\textwidth}{!}{%
\begin{tabular}{c|ccccc|ccccc|ccccc}
\toprule
\multirow{2}{*}{Batch Size} & \multicolumn{5}{c|}{Both Sides}        & \multicolumn{5}{c|}{Left Side}         & \multicolumn{5}{c}{Right Side}         \\\cmidrule{2-16}
                            & NPV   & TNR   & PPV   & TPR   & Acc   & NPV   & TNR   & PPV   & TPR   & Acc   & NPV   & TNR   & PPV   & TPR   & Acc    \\
\midrule
1                           & 0.718 & 0.700 & 0.707 & 0.725 & 0.713 & 0.737 & 0.700 & 0.714 & 0.750 & 0.725 & 0.700 & 0.700 & 0.700 & 0.700 & 0.700  \\
2                           & 0.690 & 0.725 & 0.711 & 0.675 & 0.700 & 0.700 & 0.700 & 0.700 & 0.700 & 0.700 & 0.682 & 0.750 & 0.722 & 0.650 & 0.700  \\
4                           & 0.682 & 0.750 & 0.722 & 0.650 & 0.700 & 0.682 & 0.750 & 0.722 & 0.650 & 0.700 & 0.682 & 0.750 & 0.722 & 0.650 & 0.700  \\
8                           & 0.660 & 0.775 & 0.727 & 0.600 & 0.688 & 0.652 & 0.750 & 0.706 & 0.600 & 0.675 & 0.667 & 0.800 & 0.750 & 0.600 & 0.700  \\
16                          & 0.714 & 0.750 & 0.737 & 0.700 & 0.725 & 0.714 & 0.750 & 0.737 & 0.700 & 0.725 & 0.714 & 0.750 & 0.737 & 0.700 & 0.725  \\
32                          & 0.714 & 0.750 & 0.737 & 0.700 & 0.725 & 0.714 & 0.750 & 0.737 & 0.700 & 0.725 & 0.714 & 0.750 & 0.737 & 0.700 & 0.725  \\
64                          & 0.705 & 0.775 & 0.750 & 0.675 & 0.725 & 0.714 & 0.750 & 0.737 & 0.700 & 0.725 & 0.696 & 0.800 & 0.765 & 0.650 & 0.725  \\
128                         & 0.682 & 0.750 & 0.722 & 0.650 & 0.700 & 0.714 & 0.750 & 0.737 & 0.700 & 0.725 & 0.652 & 0.750 & 0.706 & 0.600 & 0.675 \\
\bottomrule
\end{tabular}
}
\end{table*}

\begin{table*}[h]
\centering
\caption{Parameter study of SMeta-AE on different learning steps.}
\label{Ablation_learning_step}
\resizebox{\textwidth}{!}{%
\begin{tabular}{c|ccccc|ccccc|ccccc} 
\toprule
\multirow{2}{*}{\begin{tabular}[c]{@{}c@{}}\\Learning Step\end{tabular}} & \multicolumn{5}{c|}{Both Sides}                           & \multicolumn{5}{c|}{Left Side}                            & \multicolumn{5}{c}{Right Side}         \\ 
\cmidrule{2-16}
                                                                         & NPV   & TNR   & PPV   & TPR   & Acc                       & NPV   & TNR   & PPV   & TPR   & Acc                       & NPV   & TNR   & PPV   & TPR   & Acc    \\ 
\midrule
1                                                                        & 0.714 & 0.750 & 0.737 & 0.700 & 0.725                     & 0.750 & 0.750 & 0.750 & 0.750 & 0.750                     & 0.682 & 0.750 & 0.722 & 0.650 & 0.700  \\
2                                                                        & 0.711 & 0.675 & 0.690 & 0.725 & 0.700                     & 0.722 & 0.650 & 0.682 & 0.750 & 0.700                     & 0.700 & 0.700 & 0.700 & 0.700 & 0.700  \\
3                                                                        & 0.725 & 0.725 & 0.725 & 0.725 & 0.725                     & 0.750 & 0.750 & 0.750 & 0.750 & 0.750                     & 0.700 & 0.700 & 0.700 & 0.700 & 0.700  \\
4                                                                        & 0.698 & 0.750 & 0.730 & 0.675 & 0.713                     & 0.714 & 0.750 & 0.737 & 0.700 & 0.725                     & 0.682 & 0.750 & 0.722 & 0.650 & 0.700  \\
5                                                                        & 0.714 & 0.750 & 0.737 & 0.700 & 0.725                     & 0.714 & 0.750 & 0.737 & 0.700 & 0.725                     & 0.714 & 0.750 & 0.737 & 0.700 & 0.725  \\
6                                                                        & 0.667 & 0.800 & 0.750 & 0.600 & 0.700                     & 0.682 & 0.750 & 0.722 & 0.650 & 0.700                     & 0.654 & 0.850 & 0.786 & 0.550 & 0.700  \\
7                                                                        & 0.674 & 0.775 & 0.735 & 0.625 & 0.700                     & 0.682 & 0.750 & 0.722 & 0.650 & 0.700                     & 0.667 & 0.800 & 0.750 & 0.600 & 0.700  \\
8                                                                        & 0.682 & 0.750 & 0.722 & 0.650 & 0.700                     & 0.714 & 0.750 & 0.737 & 0.700 & 0.725                     & 0.652 & 0.750 & 0.706 & 0.600 & 0.675  \\                                           9        & 0.690 & 0.725 & 0.711 & 0.675 & 0.700 & 0.667 & 0.700 & 0.684 & 0.650 & 0.675 & 0.714 & 0.750 & 0.737 & 0.700 & 0.725  \\
10                                                   & 0.653 & 0.800 & 0.742 & 0.575 & 0.688 & 0.682 & 0.750 & 0.722 & 0.650 & 0.700 & 0.630 & 0.850 & 0.769 & 0.500 & 0.675  \\
\bottomrule
\end{tabular}
}
\end{table*}

\begin{table*}[h]
\centering
\caption{Parameter study of SMeta-AE on different meta-learning rate ($\alpha$).}
\label{Ablation_meta_lr}
\resizebox{\textwidth}{!}{%
\begin{tabular}{c|ccccc|ccccc|ccccc} 
\toprule
\multirow{2}{*}{\begin{tabular}[c]{@{}c@{}}\\$\alpha$\end{tabular}} & \multicolumn{5}{c|}{Both Sides}       & \multicolumn{5}{c|}{Left Side}        & \multicolumn{5}{c}{Right Side}         \\ 
\cmidrule{2-16}
                                                                      & NPV   & TNR   & PPV   & TPR   & Acc   & NPV   & TNR   & PPV   & TPR   & Acc   & NPV   & TNR   & PPV   & TPR   & Acc    \\ 
\midrule
6.0E-04                                                               & 0.698 & 0.750 & 0.730 & 0.675 & 0.713 & 0.714 & 0.750 & 0.737 & 0.700 & 0.725 & 0.682 & 0.750 & 0.722 & 0.650 & 0.700  \\
7.0E-04                                                               & 0.698 & 0.750 & 0.730 & 0.675 & 0.713 & 0.714 & 0.750 & 0.737 & 0.700 & 0.725 & 0.682 & 0.750 & 0.722 & 0.650 & 0.700  \\
8.0E-04                                                               & 0.698 & 0.750 & 0.730 & 0.675 & 0.713 & 0.714 & 0.750 & 0.737 & 0.700 & 0.725 & 0.682 & 0.750 & 0.722 & 0.650 & 0.700  \\
9.0E-04                                                               & 0.682 & 0.750 & 0.722 & 0.650 & 0.700 & 0.714 & 0.750 & 0.737 & 0.700 & 0.725 & 0.652 & 0.750 & 0.706 & 0.600 & 0.675  \\
1.0E-03                                                               & 0.714 & 0.750 & 0.737 & 0.700 & 0.725 & 0.714 & 0.750 & 0.737 & 0.700 & 0.725 & 0.714 & 0.750 & 0.737 & 0.700 & 0.725  \\
1.1E-03                                                               & 0.674 & 0.775 & 0.735 & 0.625 & 0.700 & 0.682 & 0.750 & 0.722 & 0.650 & 0.700 & 0.667 & 0.800 & 0.750 & 0.600 & 0.700  \\
1.2E-03                                                               & 0.698 & 0.750 & 0.730 & 0.675 & 0.713 & 0.682 & 0.750 & 0.722 & 0.650 & 0.700 & 0.714 & 0.750 & 0.737 & 0.700 & 0.725  \\
1.3E-03                                                               & 0.698 & 0.750 & 0.730 & 0.675 & 0.713 & 0.714 & 0.750 & 0.737 & 0.700 & 0.725 & 0.682 & 0.750 & 0.722 & 0.650 & 0.700  \\
1.4E-03                                                               & 0.707 & 0.725 & 0.718 & 0.700 & 0.713 & 0.700 & 0.700 & 0.700 & 0.700 & 0.700 & 0.714 & 0.750 & 0.737 & 0.700 & 0.725  \\
1.5E-03                                                               & 0.698 & 0.750 & 0.730 & 0.675 & 0.713 & 0.714 & 0.750 & 0.737 & 0.700 & 0.725 & 0.682 & 0.750 & 0.722 & 0.650 & 0.700  \\
\bottomrule
\end{tabular}
}
\end{table*}

\begin{table*}[h]
\centering
\caption{Parameter study of SMeta-AE on different learning rate ($\beta$).}
\label{Ablation_lr}
\resizebox{\textwidth}{!}{%
\begin{tabular}{c|ccccc|ccccc|ccccc} 
\toprule
\multirow{2}{*}{\begin{tabular}[c]{@{}c@{}}\\$\beta$\end{tabular}} & \multicolumn{5}{c|}{Both Sides}       & \multicolumn{5}{c|}{Left Side}        & \multicolumn{5}{c}{Right Side}         \\ 
\cmidrule{2-16}
                                                                      & NPV   & TNR   & PPV   & TPR   & Acc   & NPV   & TNR   & PPV   & TPR   & Acc   & NPV   & TNR   & PPV   & TPR   & Acc    \\ 
\midrule
6.0E-04                                                               & 0.698 & 0.750 & 0.730 & 0.675 & 0.713 & 0.714 & 0.750 & 0.737 & 0.700 & 0.725 & 0.682 & 0.750 & 0.722 & 0.650 & 0.700  \\
7.0E-04                                                               & 0.689 & 0.775 & 0.743 & 0.650 & 0.713 & 0.714 & 0.750 & 0.737 & 0.700 & 0.725 & 0.667 & 0.800 & 0.750 & 0.600 & 0.700  \\
8.0E-04                                                               & 0.690 & 0.725 & 0.711 & 0.675 & 0.700 & 0.737 & 0.700 & 0.714 & 0.750 & 0.725 & 0.652 & 0.750 & 0.706 & 0.600 & 0.675  \\
9.0E-04                                                               & 0.682 & 0.750 & 0.722 & 0.650 & 0.700 & 0.682 & 0.750 & 0.722 & 0.650 & 0.700 & 0.682 & 0.750 & 0.722 & 0.650 & 0.700  \\
1.0E-03                                                               & 0.714 & 0.750 & 0.737 & 0.700 & 0.725 & 0.714 & 0.750 & 0.737 & 0.700 & 0.725 & 0.714 & 0.750 & 0.737 & 0.700 & 0.725  \\
1.1E-03                                                               & 0.690 & 0.725 & 0.711 & 0.675 & 0.700 & 0.700 & 0.700 & 0.700 & 0.700 & 0.700 & 0.682 & 0.750 & 0.722 & 0.650 & 0.700  \\
1.2E-03                                                               & 0.682 & 0.750 & 0.722 & 0.650 & 0.700 & 0.682 & 0.750 & 0.722 & 0.650 & 0.700 & 0.682 & 0.750 & 0.722 & 0.650 & 0.700  \\
1.3E-03                                                               & 0.690 & 0.725 & 0.711 & 0.675 & 0.700 & 0.700 & 0.700 & 0.700 & 0.700 & 0.700 & 0.682 & 0.750 & 0.722 & 0.650 & 0.700  \\
1.4E-03                                                               & 0.698 & 0.750 & 0.730 & 0.675 & 0.713 & 0.714 & 0.750 & 0.737 & 0.700 & 0.725 & 0.682 & 0.750 & 0.722 & 0.650 & 0.700  \\
1.5E-03                                                               & 0.682 & 0.750 & 0.722 & 0.650 & 0.700 & 0.667 & 0.700 & 0.684 & 0.650 & 0.675 & 0.696 & 0.800 & 0.765 & 0.650 & 0.725  \\
\bottomrule
\end{tabular}
}
\end{table*}
\subsubsection{Hyper-parameter study} We use the parameters in Section \ref{exp_set} as the default setting and fix the random seed as 200 to test the influence of different hyper-parameters on SMeta-AE. We show the ablation study of batch size $b$, learning steps for meta-training, meta-learning rate $\alpha$, learning rate $\beta$ in Table~\ref{Ablation_batch_size}, Table~\ref{Ablation_learning_step}, Table~\ref{Ablation_meta_lr}, and Table~\ref{Ablation_lr}, respectively. We can observe that our model is stable over most parameters. Only a few parameters (e.g., $b$=128 and $\alpha$=9e-4) will slightly impair the model performance, which may be caused by the bad initialization of the random seed, but the model performance is still competitive.

\begin{figure}[h]
\centering 
\begin{subfigure}{0.24\textwidth}
  \includegraphics[width=\textwidth]{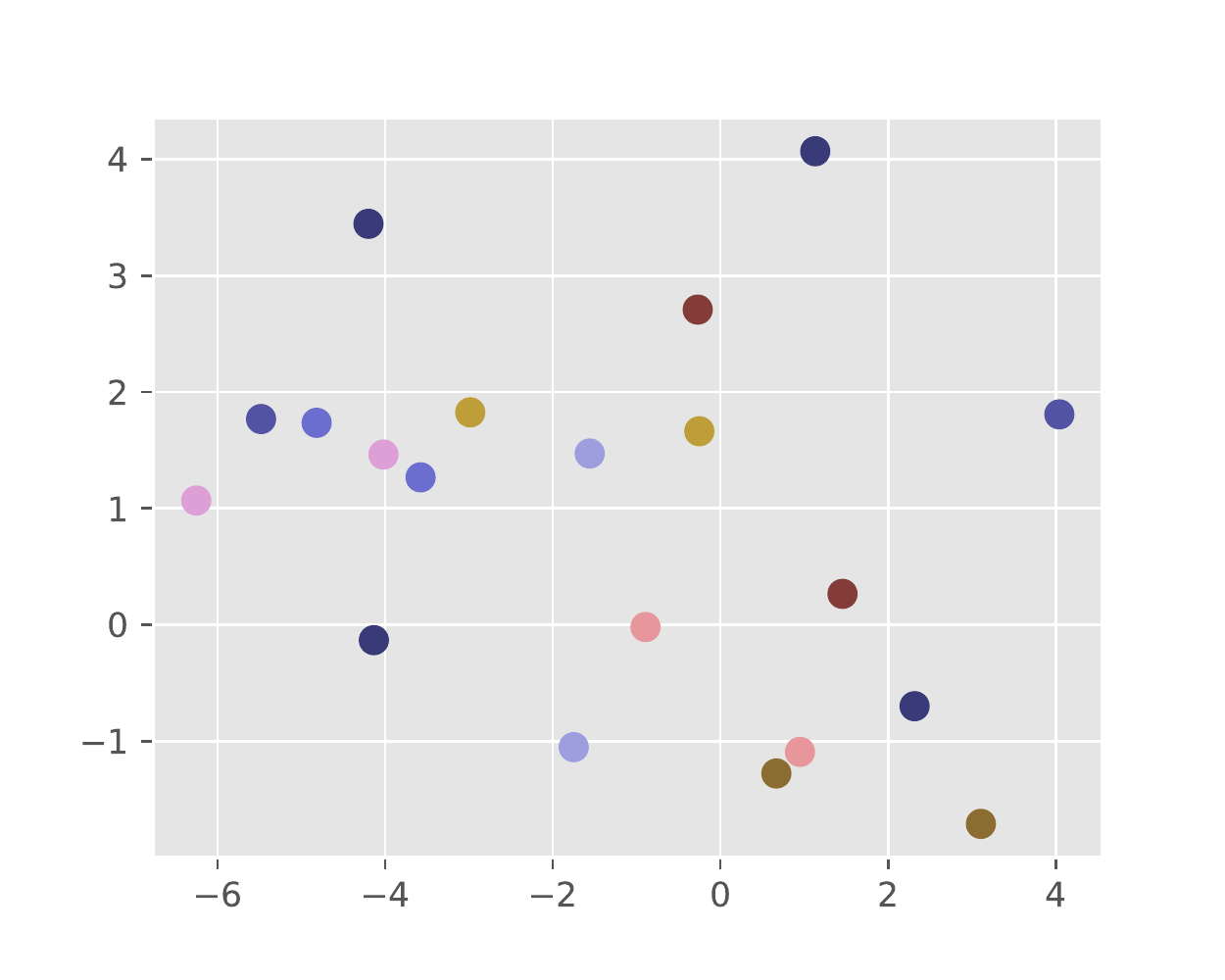}
    \centering
  \caption{Raw signals.}
\end{subfigure}\hfil 
\begin{subfigure}{0.24\textwidth}
  \includegraphics[width=\textwidth]{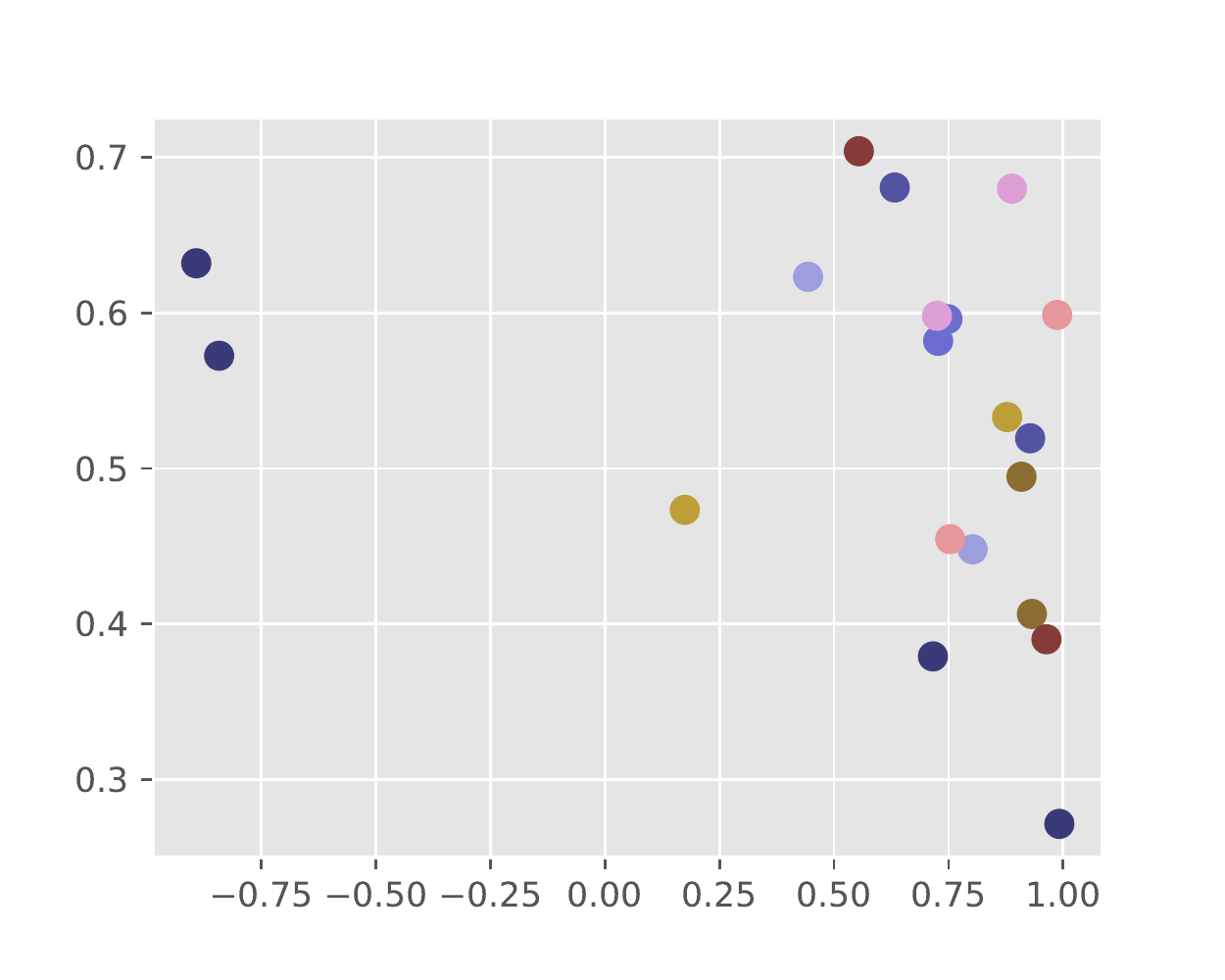}
    \centering
  \caption{Learned features.}
\end{subfigure}\hfil 
    \caption{Visualization of (a) raw signals and (b) the learned features of SMeta-SAE, respectively. We take 10 subjects from the target dataset for exhibition. Each subject (distinguished by different colors) has two signals (represented as spots).}\label{embedding}
\end{figure}
\subsubsection{Visualization of Learned Features} We take 10 subjects from the target dataset and visualize the learned features of SMeta-SAE based on t-SNE~\cite{van2008visualizing} in Figs.~\ref{embedding} (b). The samples of the same subject are plotted with the same color. Compared to the raw signals in Figs.~\ref{embedding} (a), the learned features can effectively decrease the distance between samples of the same subject, which shows the effectiveness of SMeta-learning in improving the feature qualities. The models can effectively cluster new signals if they belong to the same subjects without any prior collected data from these subjects.

\subsection{Conclusion}
We propose a novel side-aware meta-autoencoder to conduct tinnitus diagnosis across datasets for different ages and genders. We design a sliding window with down-sampling to align data, a subject-specific task sampling way for our side-aware meta-learning, and an inference stage to fit side information. We conduct extensive experiments to analyze the model performance on both ears and single ears, which is superior to conventional meta-learning. The experimental results indicate that the side information of the tested ear is beneficial for model training. Moreover, we visualize the embedding space to show the effectiveness of our method in improving the quality of the learned features. These results indicate that it is feasible to transfer model classification ability across datasets with different data collection processes. 
In the future, we plan to validate our method on more challenging datasets that collect data from more divergent experimental environments.




%

\bibliographystyle{IEEEtran}
\bibliography{bio}

\end{document}